\documentclass[twocolumn]{article}
\usepackage{graphicx}
\usepackage{amsmath}

\input{tcilatex}

\begin{document}

\title{A mathematical description of natural shapes in our nonlinear world}
\author{Ricardo Chac\'{o}n \\
Departamento de Electr\'{o}nica e Ingenier\'{\i}a Electromec\'{a}nica,\\
Escuela de Ingenier\'{\i}as Industriales, Universidad de Extremadura,\\
Apartado Postal 382, E-06071 Badajoz, Spain}
\maketitle

\begin{abstract}
Concerning the problem of shape and pattern description, a compact formula
(Gielis' formula$^{1-3}$) has recently been proposed that generates a vast
diversity of natural shapes. However, this formula is a modified version of
the equation for the circle, so that it is expressed in terms of
trigonometric functions which are inherent to \textit{linear} phenomena but
rarely appear in the description of nonlinear phenomena. In consequence, it
generates highly idealized (Platonic) forms rather than real-world forms.
Since natural shapes and patterns generally appear as a result of nonlinear
dynamical processes$^{4-7}$, it would be natural that formulas which aim to
describe them should be expressed in terms of related nonlinear functions.
Here, two examples of simple mathematical formulas which are natural
nonlinear modifications (one being a generalization) of Gielis' formula are
discussed. These formulas involve a comparable number of parameters and
provide non-Platonic representations of a vast diversity of natural shapes
and patterns by incorporating diverse aspects of asymmetry and seeming
disorder which are absent in the original Gielis' formula. It is also shown
how diverse sequences resembling some natural-world pattern evolutions are
also generated by such nonlinear formulas.---
\end{abstract}

Although nonlinearity is an ubiquitous feature of real-world phenomena, some
aspects of this fundamental property remain surprisingly poorly explored,
partly because they are rarely incorporated into the mathematical tools and
models which aim to describe even the simplest of such phenomena. In the
context of the description of natural forms, such as biological shapes, a
recent example is provided by Gielis' formula$^{1}$: 
\begin{equation}
r\left( \theta \right) =\frac{1}{\sqrt[n_{1}]{\left\vert \frac{\cos \left(
c\theta /4\right) }{a}\right\vert ^{n_{2}}+\left\vert \frac{\sin \left(
c\theta /4\right) }{b}\right\vert ^{n_{3}}}},
\end{equation}%
where $\left( r,\theta \right) $ and $(a,b,c,n_{1},n_{2},n_{3})$ are polar
coordinates and parameters, respectively. While formula (1) generates a
certain variety of abstract symmetric shapes, which in some cases are
reminiscent of natural forms, it is clear that its \textit{linear} origin
prevents it encompassing most natural forms, these being often characterized
by diverse types of \textit{asymmetry} and different degree of apparent 
\textit{disorder}. Clearly, this is because Gielis' formula essentially
describes modified circles, and hence the use of harmonic functions in its
formulation. Perhaps the simplest nonlinear generalizations of formula (1)
would be based on the substitution of the trigonometric functions by
Jacobian elliptic functions$^{8}$ (JEF's). This choice is quite natural
since the solutions of the most ubiquitous nonlinear integrable oscillators$%
^{9}$ (such as the pendulum, the Duffing, or the Helmholtz) are naturally
expressed in terms of JEF's. This is also the case for the most universal
nonlinear integrable partial differential equations,$^{10,11}$ such as the
sine-Gordon, the Korteweg-de Vries, the nonlinear Schr\"{o}dinger, or the
nonlinear heat equation.

Here the following two nonlinear (elliptic) modifications of Gielis' formula
are proposed: 
\begin{equation}
r_{I,II}\left( \theta \right) =\frac{1}{\sqrt[n_{1}]{\left\vert \frac{\cos %
\left[ \Phi _{I,II}\left( \theta \right) \right] }{a}\right\vert
^{n_{2}}+\left\vert \frac{\sin \left[ \Theta _{I,II}\left( \theta \right) %
\right] }{b}\right\vert ^{n_{3}}}},
\end{equation}%
\begin{equation}
\Phi _{I}\left( \theta \right) \equiv \func{am}\left[ \frac{K(m)}{2\pi }%
\left( c\theta +\varphi \right) ;m\right] ,
\end{equation}%
\begin{equation}
\Theta _{I}\left( \theta \right) \equiv \func{am}\left[ \frac{K(m)}{2\pi }%
\left( c\theta +\varphi ^{\prime }\right) ;m\right] ,
\end{equation}%
\begin{equation}
\Phi _{II}\left( \theta \right) \equiv \func{am}\left[ \frac{1}{2\pi }%
K\left( \frac{\lambda \theta }{2\pi }\right) \left( c\theta +\varphi \right)
;\frac{\theta }{2\pi }\right] ,
\end{equation}%
\begin{equation}
\Theta _{II}\left( \theta \right) \equiv \func{am}\left[ \frac{1}{2\pi }%
K\left( \frac{\lambda ^{\prime }\theta }{2\pi }\right) \left( c\theta
+\varphi ^{\prime }\right) ;\frac{\theta }{2\pi }\right] ,
\end{equation}%
where $\func{am}\left( u;m\right) $ is the JEF of parameter $m$, $K(m)$ is
the complete elliptic integral of the first kind, and $\left( \varphi
,\varphi ^{\prime },\lambda ,\lambda ^{\prime }\right) $ are additional
parameters. Note that $r_{I}(\theta )$ reduces to $r(\theta )$ for $%
m=\varphi =\varphi ^{\prime }=0$, while $r_{II}\left( \theta \right) $ only
coincides with $r\left( \theta \right) $ at $\theta =0$ for $\lambda
=\lambda ^{\prime }=\varphi =\varphi ^{\prime }=0$. An initial shape is
generated over the $\theta $ interval $\left[ 0,2\pi \right] $. The
dependence of the elliptic parameter on the angle coordinate in $%
r_{II}(\theta )$ is the key feature providing diverse variations of a given
initial shape. The rate and nature of such variations on an initial theme
(pattern) can be controlled by changing the parameters. Thus, one can obtain
sequences that mimic transformations of biological shapes, including growth
processes, and evolutions of vortex boundaries in fluids$^{12}$, \textit{%
inter al}. The top row of Fig. 1 shows an illustrative example of this
property for the case of a schematic (daisy-like) flower whose initial
ellipse-like petals transform into triangle-like ones. It is worth
mentioning that even such simple representations of flowers cannot be
generated by formula (1). To achieve that goal, $r(\theta )$ would have to
be multiplied by an additional harmonic function$^{1}$. The sequence was
generated using the type II formula at the parameter values $%
(a,b,c,n_{1},n_{2},n_{3},\varphi ,\varphi ^{\prime },\lambda ,\lambda
^{\prime })=\left( 2,173,12,-2,2,2,0,\pi ,1,1\right) $ over the $\theta $
intervals $\left[ 0,2\pi \right] $, $\left[ 3\pi ,5\pi \right] $, $\left[
20\pi ,22\pi \right] $, and $\left[ 9220\pi ,9222\pi \right] $,
respectively. Another important feature of real-world shapes that is
incorporated by the above nonlinear formulas is a certain disordered-looking
impression, such as that in a rose after a spring shower. This feature is
illustrated in the second row of Fig. 1, where different versions of a
rose-like flower are depicted. The first is generated using Gielis' formula
at the parameter values $(a,b,c,n_{1},n_{2},n_{3})=\left(
2,5,1.61803,5,5,5\right) $, and would represent an idealized symmetric rose.
This contrasts with the second and third versions, which were generated
using the type II formula at the additional parameters $\left( \varphi
,\varphi ^{\prime },\lambda ,\lambda ^{\prime }\right) =\left( 0,\pi
,1,1\right) $ and $\left( 0,0,0.995,0.995\right) $, respectively, and with
the fourth version which was generated using the type I formula at the
additional parameter values $(\varphi ,\varphi ^{\prime },m)=(0,\pi ,0.999)$%
. The third row of Fig. 1 shows a sequence of starfish-like patterns to
illustrate the flexibility of the above nonlinear formulas in reproducing
different aspects of asymmetry. Gielis' formula generated the perfectly
symmetric first version at the parameter values $(a,b,c,n_{1},n_{2},n_{3})=%
\left( 2,5,10,3,3,3\right) $. The type II formula generated the second and
third versions at the parameter values $(a,b,c,n_{1},n_{2},n_{3},\varphi
,\varphi ^{\prime },\lambda ,\lambda ^{\prime })=\left( 2,5,10,3,3,3,0,\pi
,1,1\right) $ and ($2$, $10$, $10$, $-2$, $2$, $2$, $0$, $\pi $, $0.91$, $1)$%
, respectively, and the type I formula generated the fourth version at the
parameter values $(a,b,c,n_{1},n_{2},n_{3},\varphi ,\varphi ^{\prime
},m)=\left( 2,10,10,-2,2,2,0,\pi ,1-10^{-14}\right) $. Formulas types I and
II are examples of an elliptic class that should certainly be useful in
graphics and modeling where several basic shapes may combine to yield more
complex pictures. Figure 2 shows a flower obtained by superposition of two
type I shapes generated at parameter values $(a,b,c,n_{1},n_{2},n_{3},%
\varphi ,\varphi ^{\prime },m)=$($2,3,10,3,3,3,0,\pi ,0.95)$ and ($%
2,17,10,3,3,3,0,\pi ,0.999)$.

In conclusion, since the price of the strikingly broad and complex diversity
of natural shapes is the inherent nonlinear character of the processes
giving rise to them, one should expect that even the most elemental formulas
aimed at describing them should be expressed in terms of nonlinear
mathematical functions closely related to general (i.e., nonlinear) natural
processes. The formulas presented here are solely intended as examples of
this line of thinking. The challenge now is, as in the case of Gielis'
formula, to relate them (or similar formulas) to some basic natural
mechanism of self-organization and pattern formation.

\paragraph{\protect\bigskip }

\subsubsection{Figure Captions}

\bigskip 

Figure 1. Shapes generated using nonlinear formulas (2)-(6).

\bigskip 

Figure 2. Flower composed of two shapes generated by formula type I.

\end{document}